\documentclass[twocolumn,prl,amsmath,amssymb,showpacs,superscriptaddress]{revtex4}
\usepackage{graphicx}
\pdfoutput=1

\begin{document}
\title{Cooper-Pair Injection into Quantum Spin Hall Insulators}
\author{Koji Sato}
\affiliation{Department of Physics and Astronomy, University of California, Los Angeles, California 90095, USA}
\author{Daniel Loss}
\affiliation{Department of Physics, University of Basel, Klingelbergstrasse 82, CH-4056 Basel, Switzerland}
\author{Yaroslav Tserkovnyak}
\affiliation{Department of Physics and Astronomy, University of California, Los Angeles, California 90095, USA}
 
\begin{abstract}
We theoretically study tunneling of Cooper pairs from a superconductor spanning a two-dimensional topological insulator strip into its helical edge states. The coherent low-energy electron-pair tunneling sets off positive current cross correlations along the edges, which reflect an interplay of two quantum-entanglement processes. Most importantly, superconducting spin pairing dictates a Cooper pair partitioning into the helical edge liquids, which transport electrons in opposite directions for opposite spin orientations. At the same time, Luttinger-liquid correlations fractionalize electrons injected at a given edge into counter-propagating charge pulses carrying definite fractions of the elementary electron charge.
\end{abstract}

\pacs{71.10.Pm,72.25.Hg,73.43.-f}


\maketitle

A quest for solid-state medium providing means to produce, transport, manipulate, and detect quantum entanglement has been fueling tremendous research activity in recent years. Many of the practical proposals rely upon an innately quantum-mechanical electron spin as an elementary building block for quantum bits and ultimately quantum computation \cite{lossPRA98}. The promise for spin-based quantum manipulations hinges on its relatively long relaxation and coherence times, in contrast to charge-based degrees of freedom that are naturally more susceptible to various decoherence processes inherent to a solid-state environment. An alternative fault-tolerant charge-based quantum computation relies on topological protection of braiding statistics of non-Abelian anyons in certain quantum Hall states \cite{nayakRMP08}, although the experimental situation there still offers a formidable challenge \cite{dolevNAT08}. In this Letter, we propose to utilize spin-dependent topological properties of newly-discovered two-dimensional quantum spin Hall insulators (QSHI) \cite{kanePRL05to,konigSCI07,qiPT10} in combination with spin entanglement of Cooper pairs (CP's) in superconductors as a starting point for spin-based macroscopic quantum manipulations \cite{burkardPRB00,recherPRB01,lesovikEPJB01,recherPRB02,benaPRL02,kimPRL04}. Recent experimental strides in realizing mesoscopic CP splitters \cite{hofstetterNAT09,herrmannPRL10} and measuring their cross correlations \cite{weiCM09} show the feasibility of our proposal.

While spins appear to be good candidates for initialization (e.g., by spin-exchange or superconducting correlations) and local gating (both by magnetic and electric fields) of quantum bits, the transport and detection remain more challenging. For the latter, some kind of spin-to-charge conversion appears to be necessary at present \cite{hansonRMP07}. In this Letter, we utilize topological helicity of the QSHI edge states to convert spin entanglement into measurable charge-current correlations, which are amenable to modern experimental capabilities \cite{blanterPRP00}. While we will not study here the feasibility of using the same system to transport spin entanglement, its potential to this end should also be clear from our discussion. Our proposal exploits the ideas \cite{recherPRB02,benaPRL02,kimPRL04} for injecting CP's into two Luttinger-liquid (LL) wires. A key role will be played by electron-electron interactions along the edges away from the contacts with the superconductor (SC), which govern nonchiral LL charge fractionalization and interactions with Fermi-liquid contacts \cite{safiPRB95} as well as suppress same-edge tunneling \cite{recherPRB02,benaPRL02}. The tunneling cross correlations thus contain a wealth of information about LL effects, SC spin pairing (especially when dealing with unconventional SC's \cite{sigristRMP91}), and topological properties of the QSHI.

\begin{figure}[t]
\includegraphics[width=0.9\linewidth]{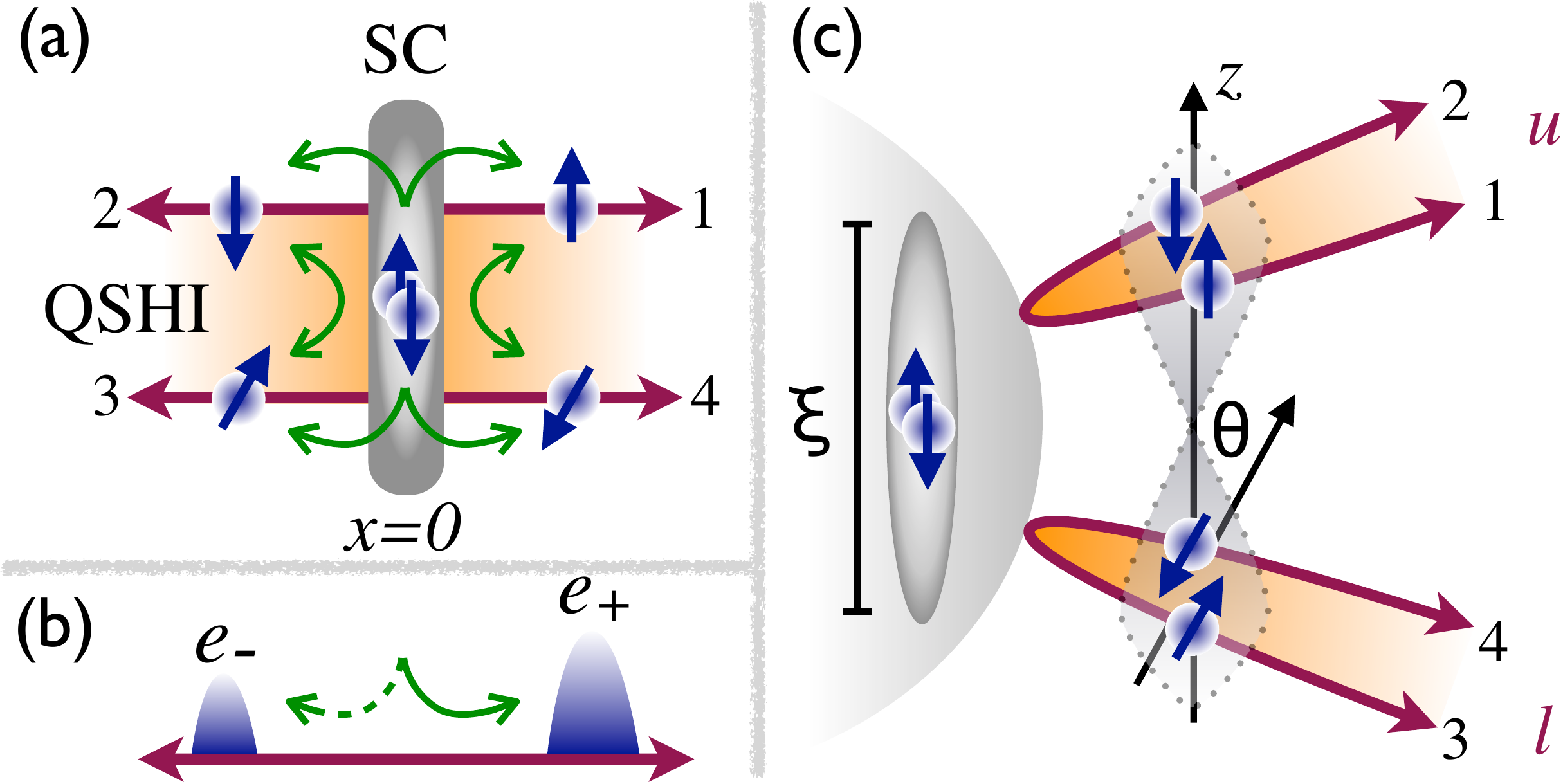}
\caption{(Color online) Schematic of the model. (a) Same-edge ($\bar{H}'$) and cross-edge ($\check{H}'$) singlet electron-pair  injection processes from the superconductor into the upper and lower QSHI edges. (b) A single right-moving electron injected into an edge LL ``pumps" a charge of $e(1-g)/2$ to the left. The net fractionalized charges propagating to the right (left) are thus respectively given by $e_\pm=e(1\pm g)/2$ \cite{safiPRB95}. (c) A schematic bending of the edges shows a close analogy of our model to the proposed entangled CP injection into carbon nanotubes \cite{benaPRL02}. The essential differences stem from the LL correlations along the edges vs those at carbon nanotubes' ends and the natural spin filtering provided by the edge-state helicity in the QSHI.}
\label{fig}
\end{figure}

At sufficiently low voltages and temperatures (on the scale set by the superconducting gap $\Delta$), two electrons that initially constitute a CP on the superconducting side are injected into the QSHI coherently, with a time delay of up to $\sim\Delta^{-1}$. The larger this delay, the weaker the LL effects on the suppression of the tunneling into the same edge \cite{recherPRB02}. Since this does not affect the low-energy tunneling exponent, however, we can use a simplified model of equal-time CP injection \cite{benaPRL02}, reinserting the delay effects \cite{recherPRB02} via the appropriate energy normalization by $\Delta$ when necessary. The (equal-time) tunneling Hamiltonian has two pieces corresponding to the CP injection into the same ($\bar{H}'$) or different ($\check{H}'$) edges at $x=0$:
\begin{align}
\label{H12b}
\bar{H}'&=\bar{\Gamma}e^{-i2eVt}\left[\psi^{(u)}_\uparrow\psi^{(u)}_\downarrow(0)+\psi^{(l)}_\uparrow\psi^{(l)}_\downarrow(0)\right]+{\rm H.c.}\,,\\
\check{H}'&=\check{\Gamma}e^{-i2eVt}\left[\psi^{(u)}_\uparrow\psi^{(l)}_\downarrow(0)-\psi^{(u)}_\downarrow\psi^{(l)}_\uparrow(0)\right]+{\rm H.c.}\,,
\label{H12v}
\end{align}
where $V$ is the voltage applied between the SC and QSHI and  $u/l$ label electrons in the upper/lower edges. $\psi_{\uparrow,\downarrow}^{(u,l)}(x)$ are electronic field operators for the helical edge modes. Figure~\ref{fig}(a) shows a schematic of our setup. Since the CP density of states in the SC is a delta function at the Fermi level, the effective tunneling Hamiltonian above determines the total current injected from the SC. We will focus on the weak-tunneling limit, where the overlap between different CP's injected in the QSHI is negligible and the spin-singlet entanglement is maximized.

In Eq.~(\ref{H12b}), we assumed symmetric tunneling into the two edges and in the second equation, we accounted for singlet pairing of the CP's. It should be clear that any asymmetry in the two tunneling contacts would enhance the relative role of the same-edge tunneling. The electron-electron repulsion, on the other hand, can reverse this tendency toward the different-edge tunneling \cite{recherPRB01,recherPRB02,benaPRL02}. There are at least two factors, however, that may hinder this most interesting scenario. First of all, the cross-edge tunneling amplitude $\check{\Gamma}$ is suppressed exponentially for the edge separations $d$ larger than the superconducting coherence length $\xi$. Even more problematically, CP injection into different edges is suppressed as a power law in $k_Fd$ (where $k_F$ is the SC's Fermi wave number) due to destructive Friedel-like interference for electrons tunneling from the SC into different contacts \cite{recherPRB02}. This issue can be overcome by injecting CP's via an intermediate layer of material with a longer Fermi wavelength and proximity-induced SC gap \cite{hofstetterNAT09}.

Electron tunneling into different edges is spin symmetric if the full structure is inversion symmetric. The latter can be affected by, e.g., applying a local strain or gate voltage to one of the edges, thus locally modifying its spin-orbit coupling. We will account for this tunable asymmetry by a relative angle $\theta$ between spin-quantization axes at the two contacts. See Figs.~\ref{fig}(a) and \ref{fig}(b). Tunneling Hamiltonian (\ref{H12v}) then has to be modified by a relative spin rotation at, say, the lower edge:
\begin{equation}
\psi_{\uparrow,\downarrow}^{(l)}\to\cos(\theta/2)\psi_{\uparrow,\downarrow}^{(l)}\pm\sin(\theta/2)\psi_{\downarrow,\uparrow}^{(l)}\,.
\label{theta}
\end{equation}
In comparison to Ref.~\cite{benaPRL02}, our $\theta$ in Eq.~(\ref{theta}) has a similar effect on the edge-current cross correlations as spin filtering in carbon nanotubes along axes misaligned by $\theta$. The key practical difference here is that such spin filtering is difficult and has not yet been achieved experimentally, while our proposal is based on measuring unfiltered charge currents. In other words, an analog of spin filtering is already built into helicity of the edge states.

The effective edge-state Hamiltonian for the QSHI, including the interband and intraband forward-scattering processes near the Fermi points is \cite{gimarchi,wuPRL06,stromPRL09}
\begin{equation}
H_0=v\sum_{k=u,l}\int\frac{dx}{2\pi}\left[\frac{1}{g}\left(\partial_{x}\phi^{(k)}\right)^{2}+g\left(\partial_{x}\theta^{(k)}\right)^{2}\right]\,.
\label{H0}
\end{equation}
Here, $\phi^{(k)}\equiv(\phi^{(k)}_\uparrow+\phi^{(k)}_\downarrow)/2$ and $\theta^{(k)}\equiv\pm(\phi^{(k)}_\uparrow-\phi^{(k)}_\downarrow)/2$ for $k=u/l$ are linear combinations of the ``spin-$s$" bosonic density operators $\phi^{(k)}_s$, such that the bosonization identity for the fermionic field operators along the upper edge with the spin-up/down electrons moving to the right/left is $\psi^{(u)}_s(x)\propto e^{is\phi^{(u)}_s(x)}$. Since the relationship between the spin and orbital chirality reverses at the opposite edge, $\psi^{(l)}_s(x)\propto e^{-is\phi^{(l)}_s(x)}$. In our convention, the commutation relations are $[\theta^{(k)}(x),\phi^{(k)}(x')]=(i\pi/2){\rm sign}(x-x')$. $g$ in Eq.~(\ref{H0}) parametrizes the strength of the electron-electron interactions ($g=1$ for free electrons and $0<g<1$ for repulsive interactions). Interband scattering between the right and left movers within a given edge leads to nontrivial correlation effects (i.e., $g\neq1$), making the QSHI fundamentally distinct from an integer quantum Hall insulator with chiral edge states \cite{wenIJMPB92}. In the special case of equal-strength interband and intraband scattering, $v=v_F/g$, in terms of the bare Fermi velocity $v_F$. In the general case with broken Galilean invariance, however, $g$ and $v$ are independent phenomenological parameters governed by the interplay between band-structure and correlation effects \cite{gimarchi}. A rough estimate \cite{stromPRL09} gives $g\gtrsim0.5$ for helical edge states in a HgTe quantum well \cite{konigSCI07}, corresponding to moderate interaction effects while the two-electron disorder and umklapp backscattering are still irrelevant \cite{wuPRL06}.

In order to measure current cross correlations, we place current sensors along the edges at points 1, 2, 3, and 4, as shown in Fig.~\ref{fig}(a). Alternatively, we can divert edge currents into voltage probes \cite{dolevNAT08} and measure low-frequency voltage correlations $\langle\delta V_i\delta V_j\rangle$.
Note, however, that the Fermi-liquid voltage probes would affect incoming currents, effectively masking the LL charge fractionalization at low voltages \cite{safiPRB95}. Using bosonic representation for the current operators along the edges, $I^{(k)}=vg\partial_x\theta^{(k)}/\pi$, we evaluate the symmetrized noise spectrum
\begin{equation}
S_{ij}(\omega)=S_{ji}(-\omega)=\int_{-\infty}^{\infty}dt e^{i\omega t}\left\langle\lbrace\delta I_i(t), \delta I_j(0)\rbrace\right\rangle\,,
\end{equation}
where $\delta I_i(t)\equiv I_i-\langle I_i(t)\rangle$. $i$ labels four outgoing channels: $i=1$, upper right; 2, upper left; 3, lower left; and 4, lower right branches. To this end, we use standard Keldysh formalism \cite{chamonPRB96}. Note that when calculating noise correlations to the leading (second) order in the tunneling coefficients $\Gamma_i$, we do not distinguish between the noise $\langle\delta I_i\delta I_j\rangle$ and current $\langle I_iI_j\rangle$ correlators, since the difference $\langle I_i\rangle\langle I_j\rangle$ enters at the fourth order in tunneling. The current direction is always chosen away from the SC.

Let us first discuss the most interesting regime, when a CP injected from a SC tunnels into two separate edges. This is governed by Eq.~(\ref{H12v}) and sketched by the cross-edge right- and left-moving electron pairs in Fig.~\ref{fig}(a). A finite $\theta$, furthermore, leads to correlations between leads 1 and 3 as well as 2 and 4. Initial spin-singlet entanglement is thus converted into the $\theta$-dependent cross-edge current correlations. In addition to this, there is a purely LL entanglement stemming from the injected charge fractionalization \cite{safiPRB95}, which has to be included on equal footing with the spin entanglement. This is sketched in Fig.~\ref{fig}(c): A single electron injected into a given edge sets off a counterpropagating nondispersive charge wave, which eats up a fraction $(1-g)/2$ of the elementary electron charge $e$.

Putting these superconducting and LL correlations together, we find for the low-frequency noise cross correlators $S_{14}=S_{23}=\check{S}_+$:
\begin{align}
\frac{\check{S}_+}{eV}&\propto|\check{\Gamma}|^2\left(\frac{eV}{\epsilon_F}\right)^{2\gamma}\left[\left(e_+^2+e_-^2\right)\cos^2\frac{\theta}{2}+2e_+e_-\sin^2\frac{\theta}{2}\right]\nonumber\\
&=|\check{\Gamma}|^2\left(\frac{eV}{\epsilon_F}\right)^{2\gamma}e^2\frac{1+g^2\cos\theta}{2}\,.
\label{Sp}
\end{align}
We are focusing on the low-temperature regime, $k_BT\ll eV$, so that thermal noise can be neglected. $\epsilon_F$ is the characteristic bandwidth of the LL, typically of the order of the Fermi energy, and $\gamma\equiv(g+g^{-1})/2-1$ is the spinless nonchiral LL bulk exponent for a single-electron injection \cite{gimarchi}. Apart from the generic LL tunneling anomaly, there are two types of interesting factors in Eq.~(\ref{Sp}): LL factors $e_+^2+e_-^2=e^2(1+g^2)/2$ and $2e_+e_-=e^2(1-g^2)/2$ reflecting charge fractionalization \cite{safiPRB95} and trigonometric factors $\cos^2(\theta/2)$ and $\sin^2(\theta/2)$ reflecting spin-singlet entanglement. Similarly, we find for $S_{13}=S_{24}=\check{S}_-$:
\begin{align}
\frac{\check{S}_-}{eV}&\propto|\check{\Gamma}|^2\left(\frac{eV}{\epsilon_F}\right)^{2\gamma}\left[\left(e_+^2+e_-^2\right)\sin^2\frac{\theta}{2}+2e_+e_-\cos^2\frac{\theta}{2}\right]\nonumber\\
&=|\check{\Gamma}|^2\left(\frac{eV}{\epsilon_F}\right)^{2\gamma}e^2\frac{1-g^2\cos\theta}{2}\,.
\label{Sm}
\end{align}
For the total interedge correlation $S_{ul}\equiv\langle I_uI_l\rangle=S_{13}+S_{14}+S_{23}+S_{24}=2(\check{S}_++\check{S}_-)$, we have $S_{ul}/V\propto V^{2\gamma}$, independent of the charge fractionalization and angle $\theta$. The self-correlators are determined by the respective average currents, as generally expected for the Poissonian statistics in the tunneling regime:
\begin{equation}
S_{ii}=2\left(e_+^2+e_-^2\right)\langle I_i\rangle/e=(1+g^2)e\langle I_i\rangle\,.
\label{Sii}
\end{equation}

In the special case when $\theta=0$ and $g=1$, there is a perfect positive cross correlation of currents at contacts 1, 4 and 2, 3 (i.e., $\check{S}_+=S_{ii}$) and zero correlation at contacts 1, 3 and 2, 4 (i.e., $\check{S}_-=0$). The former is certainly a manifestation of many-body entanglement, since noninteracting electrons are required to have nonpositive cross correlations in general multiterminal structures \cite{blanterPRP00}. According to Eqs.~(\ref{Sp}) and (\ref{Sm}), the LL parameter $g$ can be measured in the inversion-symmetric configuration (so that $\theta=0$) using
\begin{equation}
g=\sqrt{\left(\check{S}_+-\check{S}_-\right)/\left(\check{S}_++\check{S}_-\right)}\,.
\label{g}
\end{equation}
If $\theta$ is unknown and not precisely controlled, but can, nonetheless, be swept over the half-period $(0,\pi)$, $g$ will be found by maximizing the absolute value of the right-hand side in Eq.~(\ref{g}), which equals $g\sqrt{\cos\theta}$ according to Eqs.~(\ref{Sp}), (\ref{Sm}). The same $g$ experimentally extracted from Eqs.~(\ref{Sii}) and (\ref{g}) would provide direct evidence of spin entanglement as manifested by the interedge correlations (even if $g$ turns out to be trivial, i.e., close to 1).

Detrimental to the above formulation would be any backscattering along the edges. While the time-reversal symmetry protects against backscattering on nonmagnetic impurities, there would inevitably be interactions of the edge modes with nearby Fermi liquids (such as electrostatic gates, metallic contacts etc.), and the edges would terminate somewhere, possibly merging with a Fermi-liquid continuum. As was already mentioned above, the Fermi-liquid probes measuring current fluctuations would conceal the LL fractionalization \cite{safiPRB95}, effectively removing positive cross correlations associated with the LL physics. In this case, we would expect $g$ in Eqs.~(\ref{Sp})-(\ref{g}) to be close to unity. More problematically, backscattering engendered by charge fluctuations on nearby gates or other metallic regions, which would scramble ballistic propagation of edge states, needs to be eliminated from a practical implementation of our proposal. It should be clear, however, that various backscattering and boundary-related effects should not affect the finite-frequency correlations on time scales shorter than those at which they set in while still longer than $\Delta^{-1}$, so that the CP tunneling delay is not resolved.

In the absence of the LL fractionalization, $g=1$, the noise $\check{S}_\pm\propto(1\pm\cos\theta)/2$ has the same angular dependence as the spin-current noise in the carbon-nanotube proposal of Ref.~\cite{benaPRL02}. The $\theta$ dependence here is the same, in turn, as in the Einstein-Podolsky-Rosen thought experiment \cite{einsteinPR35} on a delocalized spin-singlet state of two electrons. It is instructive to understand a close relationship between our setup and that of Ref.~\cite{benaPRL02} by schematically folding two sides of either infinite QSHI edge into a semi-infinite ``wire," as sketched in Fig.~\ref{fig}(b). This wire is then formally identical to a semi-infinite spinful LL, only without interactions between spins up (down) moving in opposite directions as well as between spins up and down moving in the same direction. There is still, however, a nontrivial interaction between spins up and down moving in opposite directions, which corresponds to $g\neq1$ in our original model shown in Fig.~\ref{fig}(a). Hypothetical spin filtration \cite{benaPRL02} in semi-infinite carbon nanotubes is thus provided in our system by effectively unfolding the 1D modes of such semi-infinite wires into infinite edges, with half the number of 1D bands and helical spin character (i.e., spins up and down going in opposite directions).

Owing to the difference in the interaction channels here and in Ref.~\cite{benaPRL02}, we find a hierarchy of the tunneling exponents that is more similar to that discussed in Ref.~\cite{recherPRB02} for infinite LL wires. Injecting a CP into opposite QSHI edges is characterized by the exponent $2\gamma=g+g^{-1}-2$, as has already been discussed above [Eqs.~(\ref{Sp}) and (\ref{Sm})]. We find that the same-edge tunneling of a singlet CP is suppressed by an additional factor $\sim(eV/\Delta)^{2(\gamma'-\gamma)}$ with respect to Eqs.~(\ref{Sp}) and (\ref{Sm}), where $\gamma'\equiv g^{-1}-1$ is the end-tunneling exponent for a nonchiral spinless LL \cite{gimarchi}. The fact that $\gamma'>\gamma$ for $g<1$ insures that the split-tunneling dominates at low energies \cite{recherPRB02,benaPRL02} over the same-edge tunneling, which is parasitic for our purposes. In particular, the same-edge tunneling would introduce large positive cross correlations $S_{12}$ and $S_{34}$, while not contributing to the interedge correlations.

The situation is quite different in the case of unconventional SC's with triplet spin pairing \cite{sigristRMP91}. For example, a CP in the equal-spin pairing phase could be injected into the same (left- or right-moving) edge mode, in the case of the same-edge tunneling, which is governed by the LL exponent $4\gamma+2$ vs $2\gamma$ for the cross-edge tunneling. The relative suppression factor here is $\sim(eV/\Delta)^{2\gamma+2}$, with the quadratic factor stemming from the Pauli exclusion that hinders injection of two same-spin electrons into the same edge band \cite{fisherPRB94}. For a finite $\theta$, there are also nontrivial interedge correlations, dictated by the superconducting spin pairing. In this context, we can speculate, furthermore, that a QSHI not only provides a fruitful medium for injecting and subsequently manipulating spin-entangled CP's but can also serve as a probe for spin pairing, which can thus be manifested via current cross correlations along the edges. This offers an alternative to the proposal of experimentally probing the superconducting order parameter via nonlocal spin and charge pumping \cite{brataasPRL04}.

We finally note that our setup is distinct from that of the Hanbury Brown and Twiss correlations of CP's in the QSHI discussed recently in Ref.~\cite{choiCM10}. In the latter implementation, CP's are injected ballistically from an edge portion that opens a gap due to a strong proximity effect to a nearby SC. The induced superconductivity provides perfect Andreev reflection for electrons incoming from the normal region of the edge. Time-reversal symmetry prohibits normal scattering in this situation. This Andreev reflection is found to induce ordinary (i.e., negative) current cross correlations with the second edge that is connected to the first by a point contact \cite{choiCM10}. In the present situation, on the other hand, the contact between the SC and the QSHI is weak and the electrons are thus passing along the edges almost ballistically with little Andreev scattering. Those electrons that rarely undergo Andreev reflection, in turn, scatter to a different edge as holes, which is called crossed Andreev reflection (while the same-edge Andreev reflection is suppressed by the LL correlations). This effective flipping of the scattered carrier charge reconciles our positive cross correlations with the Hanbury Brown and Twiss perspective of Ref.~\cite{choiCM10}.

We are grateful to A. Brataas and S. Valenzuela for stimulating discussions and P. Recher for valuable comments on the manuscript. This work was supported by the Alfred~P. Sloan Foundation, by the NSF under Grant No.~DMR-0840965 (Y.T.), and by the Swiss NSF and DARPA QuEST (D.L.).

\end{document}